# Steep Slope Hysteresis-free Negative Capacitance MoS$_2$ Transistors

Mengwei Si[1,3], Chun-Jung Su[2], Chunsheng Jiang[1,4], Nathan J. Conrad[1,3], Hong Zhou[1,3], Kerry D. Maize[1,3], Gang Qiu[1,3], Chien-Ting Wu[2], Ali Shakouri[1,3], Muhammad A. Alam[1] and Peide D. Ye*[,1,3]

[1]*School of Electrical and Computer Engineering, Purdue University, West Lafayette, Indiana 47907, United States*

[2]*National Nano Device Laboratories, Hsinchu 300, Taiwan*

[3]*Birck Nanotechnology Center, Purdue University, West Lafayette, Indiana 47907, United States*

[4]*Tsinghua National Laboratory for Information Science and Technology, Institute of Microelectronics, Tsinghua University, Beijing 100084, China*

\* Address correspondence to: yep@purdue.edu (P.D.Y.)

The so-called Boltzmann Tyranny defines the fundamental thermionic limit of the subthreshold slope (SS) of a metal-oxide-semiconductor field-effect transistor (MOSFET) at 60 mV/dec at room temperature and, therefore, precludes the lowering of the supply voltage and the overall power consumption[1,2]. Adding a ferroelectric negative capacitor to the gate stack of a MOSFET may offer a promising solution to bypassing this fundamental barrier[3]. Meanwhile, two-dimensional (2D) semiconductors, such as atomically thin transition metal dichalcogenides (TMDs) due to their low dielectric constant, and ease of integration in a junctionless transistor topology, offer enhanced electrostatic control of the channel[4-12]. Here, we combine these two advantages and demonstrate for the first time a molybdenum disulfide (MoS$_2$) 2D steep slope transistor with a ferroelectric hafnium zirconium oxide layer (HZO) in the gate dielectric stack. This device exhibits excellent performance in *both* on- and off-states, with maximum drain current of 510 µA/µm, sub-thermionic subthreshold slope and is essentially hysteresis-free. Negative differential resistance (NDR) was observed at room temperature in the MoS$_2$ negative capacitance field-effect-transistors (NC-FETs) as the result of negative capacitance due to the negative drain-induced-barrier-lowering (DIBL). High on-current induced self-heating effect was also observed and studied.

TMDs have been intensely explored as 2D semiconductors for future device technologies. Atomically thin MoS$_2$ has been extensively studied as a highly promising channel material because it offers the ideal electrostatic control of the channel, ambient stability, an appropriate direct bandgap and moderate mobility. The TMD is generally configured in a junctionless (JL) form, where metal-semiconductor contacts replace the source/drain p-n junctions of a bulk transistor. Junctionless MoS$_2$ FETs exhibit high on/off ratio and strong immunity to short channel effects for transistor applications with channel length (L$_{ch}$) down to sub-5 nm[4-12]. However, the power dissipation issue remains unresolved as silicon-based MOSFETs scaling. To overcome the thermionic limit, several novel device concepts have been proposed with potential SS less than 60 mV/dec at room temperature such as impact-ionization FETs (II-FET)[13], tunneling FETs (T-FET)[14,15], nanoelectromechanical FETs (NEMFET)[16] and NC-FETs[17-28]. In a NC-FET, the insulating ferroelectric layer served as a negative capacitor so that channel surface potential can be amplified more than the gate voltage, and hence the device can operate with SS less than 60 mV/dec at room temperature[3]. The simultaneous fulfillment of internal gain and non-hysteretic condition is crucial to the proper design of capacitance matching in a stable NC-FET. Meanwhile, the channel transport in NC-FETs remains unperturbed. Therefore, coupled with the flatness of the body capacitance of TMD materials and symmetrical operation around the zero-charge point in a junctionless transistor, performance in 2D JL-NCFET is expected to improve for *both* on- and off-states. Therefore, it would be highly desirable to integrate ferroelectric insulator and 2D ultrathin channel materials as a 2D JL-NCFET to achieve high on-state performance for high operating speed and sub-thermionic SS for low power dissipation.

Here, we demonstrate steep slope $MoS_2$ NC-FETs by introducing ferroelectric HZO into the gate stack. These transistors exhibit essentially hysteresis-free switching characteristics with maximum drain current of 510 µA/µm and sub-thermionic subthreshold slope. The maximum drain current of the NC-FETs fabricated in this work is found to be around five times larger than $MoS_2$ FETs fabricated on 90 nm $SiO_2$ using the same process. As will be discussed below, this is a direct consequence of on-state voltage application in a JL-NCFET. Negative differential resistance, correlated to the negative DIBL at off-state, is observed because of drain coupled negative capacitance effect. Remarkably, the high performance sustains despite significant self-heating in the transistors, as opposed to traditional bulk MOSFETs.

The experimental device schematic of a MoS$_2$ NC-FET, as shown in Fig. 1a, consist of a mono-layer up to dozen layers of MoS$_2$ as channel, 2 nm amorphous aluminum oxide (Al$_2$O$_3$) layer and 20 nm polycrystalline HZO layer as the gate dielectric, heavily doped silicon substrate as the gate electrode and nickel source/drain contacts. HZO is chosen for its ferroelectricity, CMOS compatible manufacturing, and ability to scale down equivalent oxide thickness (EOT) to ultra-thin dimensions[23-28]. An amorphous Al$_2$O$_3$ layer was applied for capacitance matching and gate leakage current reduction through polycrystalline HZO. A cross-sectional transmission electron microscopy (TEM) image of a representative MoS$_2$ NC-FET is shown in Fig. 1b and detailed energy dispersive X-ray spectrometry (EDS) elemental mapping is presented in Fig. 1c. The EDS analysis confirms the presence and uniform distribution of elements Hf, Zr, Al, O, Mo and S. No obvious inter-diffusion of Hf, Zr and Al is found. A detailed measurement of the gate stack on rapid thermal annealing (RTA) temperature dependence using metal-oxide-semiconductor capacitor structure was carried out using fast I-V measurement. The measured hysteresis loops for polarization versus electric field (P-E) and XRD results suggest 400-500 °C RTA after atomic layer deposition (ALD) deposition contributes to enhance the ferroelectricity (Supplementary Section 1).

The electrical characteristics of MoS$_2$ NC-FETs are strongly dependent on the ferroelectricity of HZO layer, defined by the film annealing temperature and gate-to-source voltage (V$_{GS}$) sweep speed. In addition to standard I-V measurements, the "hysteresis" is measured as V$_{GS}$-difference between forward (from low to high) and reverse (from high to low) V$_{GS}$ sweeps at I$_D$=1 nA/μm and at V$_{DS}$=0.1 V. Here, we first study the room temperature characteristics of MoS$_2$ NC-FETs. Fig. 2a shows the I$_D$-V$_{GS}$ characteristics of a device with 500 °C annealed gate dielectric, measured at V$_{GS}$ step of 0.5 mV. This device has a channel length of 2 μm, channel width of 3.2 μm and channel thickness of 8.6 nm. The hysteresis (~12 mV) is small and essentially negligible, consistent with the theory of NC-FET. Gate leakage current ($I_G$) is negligible (Supplementary section 2). Fig. 2b shows SS vs. $I_D$ data of the same device as in Fig. 2a, and the comparison with simulation results and experimental results with 20 nm $Al_2O_3$ only as gate dielectric. The $MoS_2$ FETs fabricated on a 20 nm $Al_2O_3$ conventional dielectric present the typical SS of 80-90 mV/dec, much larger than the values from NC-FETs. SS is extracted for both forward sweep ($SS_{For}$) and reverse sweep ($SS_{Rev}$). The device exhibits $SS_{Rev}$=52.3 mV/dec, $SS_{For}$=57.6 mV/dec. SS below 60 mV/dec at room temperature is conclusively demonstrated for both forward and reverse sweeps at this near hysteresis-free device.

Since the HZO polarization depends on sweep-rate, the electrical characterization for the $MoS_2$ NC-FETs is also carried out at different $V_{GS}$ sweeping speeds. The sweeping speed is controlled by modifying the $V_{GS}$ measurement step, from 0.3 mV to 5 mV. Fig. 2c shows the $I_D$-$V_{GS}$ characteristics of a few-layer $MoS_2$ NC-FET measured at slow, medium and fast sweep speed, corresponding to $V_{GS}$=0.3, 1 and 5 mV. Hysteresis of the $MoS_2$ NC-FETs is found to be diminished by reducing the sweeping speed. A plateau and a minima characterize the SS (vs $I_D$) during reverse sweep. These features ($SS_{Rev,min\#1}$ and $SS_{Rev,min\#2}$) are observed among almost all fabricated devices when measured with fast sweep $V_{GS}$, as shown in Fig. 2d. The second local minimum of SS is the result of the switching between two polarization states of the ferroelectric oxide, associated with loss of capacitance matching at high speed. When measured in fast sweep mode where $V_{GS}$ step is 5 mV, the device exhibits $SS_{For}$=59.6 mV/dec, $SS_{Rev,min\#1}$=41.7 mV/dec, and $SS_{Rev,min\#2}$=5.6 mV/dec. Overall, average SS less than 60 mV/dec for over 4 decades of drain current. In slow sweep mode, no obvious second local minima and hysteresis can be observed as shown in Fig. 2a, reflecting well-matched capacitances throughout the subthreshold region. Fig.

2e shows the thickness dependence of SS from mono-layer to 5 layers of MoS$_2$ as channels (See supplementary section 4 for layer number determination). No obvious thickness dependence SS is observed. Fig. 2f shows the temperature dependence of SS for a MoS$_2$ NC FET measured from 280 K down to 160 K. Measured SS is below the thermionic limit down to 220 K. SS below 190 K is above the thermionic limit because of the stronger impact of Schottky barrier at lower temperatures. Detailed I-V characteristics at low temperature can be found in supplementary section 5.

Although the above MoS$_2$ NC-FET shows average SS during reverse sweep less than 60 mV/dec for more than 4 decades, low hysteresis is generally required for any transistor application. A detailed discussion on the non-hysteretic and internal gain conditions of MoS$_2$ NC-FET can be found in supplementary section 7 by using experimentally measured P-E results directly on HZO films. We find that both SS and hysteresis in MoS$_2$ NC-FETs is sensitive to the annealing temperature on gate dielectric. The dependence of SS on different RTA temperature is systematically studied (Supplementary Section 3). It is found that MoS$_2$ NC-FETs with RTA at 400 $^o$C and 500 $^o$C have smaller SS compared to as-grown samples and 600 $^o$C annealed samples, as shown in Fig. S4. This conclusion can be obtained similarly from hysteresis loop of P-E because gate stack with RTA at 400 $^o$C and 500 $^o$C show larger remnant polarization, indicating stronger ferroelectricity. A statistical study on temperature dependent hysteresis is shown in Fig. S4d. It is found that MoS$_2$ NC-FETs with 500 $^o$C RTA exhibit the lowest hysteresis comparing with devices without RTA, devices with RTA at 400 $^o$C and 600 $^o$C. Therefore, RTA temperature engineering could be useful and important to balance the request for both steep slope and low hysteresis.

Drain-induced-barrier-lowering is widely observed as one of the major evidences for the short channel effects in MOSFETs[2]. In conventional MOSFETs, the threshold voltage ($V_{th}$) shifts toward the negative direction as drain voltage. The DIBL, defined as DIBL=$-\Delta V_{th}/\Delta V_{DS}$, is usually positive. It has been theoretically predicted that with ferroelectric insulator introduced into gate stack of a practical transistor, the DIBL could be reversed in NC-FETs[29]. NDR can naturally occur as a result of the negative DIBL effect. Fig. 3a shows the negative DIBL in $I_D$-$V_{GS}$ characteristics of another device with a channel length of 2 μm, a channel width of 5.6 μm, a channel thickness of 7.1 nm and 2 nm $Al_2O_3$ and 20 nm HZO as gate dielectric. It is evident that the $I_D$-$V_{GS}$ curve shifts positively when $V_{DS}$ is increased from 0.1 V to 0.5 V. As this negative DIBL happens around off-state, NDR is also observed simultaneously in the same device at the off-state as shown in Fig. 2b. Fig. 3c shows the illustration of band diagram of negative DIBL effect. The negative DIBL origins from the capacitance coupling to from drain to interfacial layer between $Al_2O_3$ and HZO. The interfacial layer potential ($V_{mos}$) can be estimated as a constant when the thickness of ferroelectric oxide layer is thin (Supplementary section 7). Simulation of $V_{mos}$ shows when $V_{DS}$ is increased, the interfacial potential is reduced (Fig. 3d), indicating the carrier density in $MoS_2$ channel is reduced. Thus, the channel resistance is increased which lead to the NDR effect.

The EOT of the gate stack (2 nm $Al_2O_3$ and 20 nm HZO) in this work is measured to be 4.4 nm by C-V measurement. The breakdown voltage is consistently measured to be around 11 V. The breakdown voltage/EOT is 2.5 V/nm, which is about 2.5 times larger than the value of $SiO_2$. It can be easily verified that the breakdown voltage/EOT is proportional to the electric displacement field. As it is well known from Maxwell's equations that electric displacement field is proportional to the charge density, higher breakdown voltage/EOT could lead to higher carrier density. Fig. 4a shows the $I_D$-$V_{DS}$ characteristics measured at room temperature of a $MoS_2$ NC FET with 100 nm channel length. The thickness of the $MoS_2$ flake is 3 nm. The gate voltage was stressed up to 9 V and maximum gate voltage over EOT in this device is about 2 V/nm. Maximum drain current of 510 µA/µm is achieved, which is about 5 times larger than the control devices using 90 nm $SiO_2$ as gate dielectric. Note that this maximum drain current is obtained without special contact engineering such as doping[11] or heterostructure contact stack[10]; indeed, as discussed in the supplementary information, the junctionless topology is the key to improved performance of the transistor. It is an important but unexplored advantage in applying ferroelectric gate stack to enhance on-state performance. Another type of NDR (Fig. 4b) is also clearly observed when the device is biased at high $V_{GS}$ because of the self-heating effect from large drain current and voltage. Fig 4c shows the thermo-reflectance image taken at different power density from 0.6 W/mm to 1.8 W/mm. The heated channel with the increased temperature up to ~40 ºC suggests the self-heating effect, which potentially degrades channel mobility and limits the maximum drain current, has to be considered in $MoS_2$ NC-FETs.

In conclusion, we have successfully demonstrated $MoS_2$ 2D NC-FETs with the simultaneous promising on- and off-state characteristics. The stable, non-hysteretic and bi-directional sub-thermionic switching characteristics are unambiguously confirmed to be the result of negative capacitance effect. On-state performance is enhanced at the same time with a maximum drain current of 510 µA/µm at room temperature, which leads to self-heating effect. Finally, we've shown that the observed negative differential resistance is induced by the negative DIBL effect.

**Acknowledgements**

This material is based upon work partly supported by AFOSR/NSF 2DARE program, ARO, and SRC.

**Author contribution**

P.D.Y. conceived the idea and supervised the experiments. C.J.S. did the ALD of HZO and $Al_2O_3$ and dielectric physical analysis. M.S. performed the device fabrication, DC and CV measurements, and data analysis. M.S. and N.J.C. carried out the fast I-V measurement. M.S. and G.Q. did the AFM measurement. H.Z., K.D.M, and A.S. did the thermo-reflectance imaging. G.Q. performed the Raman and PL experiment. C.T.W. conducted TEM and EDS analyses. C.J. and A.M.A. conducted the theoretical calculations and analysis. M.S., A.M.A. and P.D.Y. summarized the manuscript and all authors commented on it.

**Competing financial interests statement**

The authors declare no competing financial interests.

**Figure captions**

**Figure 1 | Schematic and fabrication of MoS$_2$ NC-FETs. a** Schematic view of a MoS$_2$ NC-FET. The gate stack includes the heavily doped Si as gate electrode, 20 nm HZO as the ferroelectric capacitor, 2 nm Al$_2$O$_3$ as capping layer and capacitance matching layer. 100 nm Ni was deposited using e-beam evaporator as source/drain electrode. **b** Cross-sectional view of a representative sample showing bi-layer MoS$_2$ channel, amorphous Al$_2$O$_3$ and polycrystalline HZO gate dielectric. **c.** Corresponding EDS elemental mapping showing the distribution of elements of Hf, Zr, Al, O, Mo and S.

**Figure 2 | Off-state switching characteristics of MoS$_2$ NC-FETs. a** I$_D$-V$_{GS}$ characteristics measured at room temperature and at V$_{DS}$ from 0.1 V to 0.9 V. V$_{GS}$ step is 0.5 mV. The thickness of the MoS$_2$ flake is 8.6 nm, measured from AFM. This device has a channel length of 2 μm and channel width of 3.2 μm, RTA was performed at 500 °C during substrate preparation. **b** SS versus I$_D$ characteristics of the same device in Fig. 2a, showing minimum SS below 60 mV/dec for both forward and reverse sweep. And the comparison of SS versus I$_D$ characteristics with simulation results on the same device structure and experimental MoS$_2$ FET with 20 nm Al$_2$O$_3$ only as gate oxide. **c** I$_D$-V$_{GS}$ characteristics measured at room temperature and at V$_{DS}$=0.1 V at different gate voltage sweep speed. V$_{GS}$ step was set to be from 0.3 mV to 5 mV. The thickness of the MoS$_2$ flake is 5.1 nm. This device has a channel length of 1 μm and channel width of 1.56 μm. RTA temperature was 400 °C on gate dielectric. **d** SS versus I$_D$ characteristics during fast reverse sweep of the same device in Fig. 2c. The SS versus I$_D$ characteristics show two local minima, defined as min #1 and min #2. The min #2 suggests the switching between different polarization states of the ferroelectric HZO. **e** Layer dependence of SS from 1 layer to 5 layers. The SS of MoS$_2$ NC–FETs shows weak thickness dependence. **f** Temperature dependence of SS from 160 K layer to 280 K. Measured SS is below the thermionic limit down to 220 K. SS below 190 K shows above the thermionic limit because of stronger impact of Schottky barrier on SS.

**Figure 3 | NDR and negative DIBL in MoS$_2$ NC-FETs. a** $I_D$-$V_{GS}$ characteristics measured at room temperature and at $V_{DS}$ at 0.1 V and 0.5 V. $V_{GS}$ step during measurement was 5 mV. Inset: zoom-in image of $I_D$-$V_{GS}$ curve between -0.8 V to -0.7 V. A threshold voltage shift toward positive direction can be observed at high $V_{DS}$, indicating negative DIBL effect. The thickness of the MoS$_2$ flake is 5.3 nm, estimated from AFM characterization. This device has a channel length of 2 μm and channel width 5.6 μm. 500 °C RTA in N$_2$ for 1 min was done during substrate preparation on gate dielectric for this device. **b** $I_D$-$V_{DS}$ characteristics measured at room temperature at $V_{GS}$ from -0.65 V to -0.55 V in 0.025 V step. Clear NDR can be observed because of the negative DIBL effect induced by negative capacitance. **c**. Illustration of band diagram of negative DIBL effect. The negative DIBL origins from the capacitance coupling to from drain to interfacial layer between Al$_2$O$_3$ and HZO. **d**. Simulation of interfacial potential vs. $V_{DS}$. When $V_{DS}$ is increased, the interfacial potential is reduced so that the carrier density in MoS$_2$ channel is reduced. Thus, the channel resistance is increased and drain current is reduced.

**Figure 4 | On-state characteristics and self-heating of MoS$_2$ NC-FETs. a** $I_D$-$V_{DS}$ characteristics measured at room temperature at $V_{GS}$ from -1 V to 9 V in 0.5 V step. The thickness of the MoS$_2$ flake is 3 nm. This device has a channel length of 100 nm. The maximum stress voltage over EOT in this device is about 2 V/nm. Maximum drain current is 510 μA/μm. Clear negative drain differential resistance can be observed at high $V_{GS}$. **b** $g_D$-$V_{DS}$ characteristics from Fig. 4a at $V_{GS}$=9 V. $g_D$ less than zero at high $V_{DS}$ highlights the NDR effect due to self-heating. **c** Thermoreflectance image and **d** temperature map at different power density from 0.6 W/mm to 1.8 W/mm. The heated channel suggests that the self-heating effect has to be considered in MoS$_2$ NC-FETs with large drain current.

**Methods**

**ALD Deposition.** $Hf_{1-x}Zr_xO_2$ film was deposited on a heavily doped silicon substrate. Prior to deposition, the substrate was cleaned by RCA standard cleaning and diluted HF dip, to remove organic, metallic contaminants, particles and unintentional oxides, followed deionized water rinse and drying. The substrate was then transferred to an ALD chamber to deposit $Hf_{1-x}Zr_xO_2$ film at 250 °C, using $[(CH_3)_2N]_4Hf$ (TDMAHf), $[(CH_3)_2N]_4Zr$ (TDMAZr), and $H_2O$ as the Hf precursor, Zr precursor, and oxygen source, respectively. The $Hf_{1-x}Zr_xO_2$ film with x = 0.5 was achieved by controlling $HfO_2$:$ZrO_2$ cycle ratio of 1:1. To encapsulate the $Hf_{1-x}Zr_xO_2$ film, an $Al_2O_3$ was subsequently *in-situ* deposited using $Al(CH_3)_3$ (TMA) and $H_2O$ also at 250 °C.

**Device Fabrication.** 20 nm $Hf_{0.5}Zr_{0.5}O_2$ was deposited by ALD as a ferroelectric insulator layer on heavily doped silicon substrate after standard surface cleaning. Another 10 nm aluminum oxide layer was deposited as an encapsulation layer to prevent the degradation of HZO by the reaction with moisture in air. $BCl_3$/Ar dry etching process was carried out to adjust the thickness of $Al_2O_3$ down to 2 nm for capacitance matching. The annealing process was then performed in rapid thermal annealing in nitrogen ambient for 1 minute at various temperatures. $MoS_2$ flakes were transferred to the substrate by scotch tape-based mechanical exfoliation. Electrical contacts using 100 nm nickel electrode were fabricated using electron-beam lithography, electron-beam evaporation and lift-off process.

**Device Characterization.** The thickness of the $MoS_2$ was measured using a Veeco Dimension 3100 AFM system. DC electrical characterization was performed with a Keysight B1500 system. Fast I-V measurement was carried out using a Keysight B1530A fast measurement unit. C-V measurement was done with an Agilent E4980A LCR meter. Room temperature electrical data was collected with a Cascade Summit probe station and low temperature electrical data was collected with a Lakeshore TTP4 probe station. Thermoreflectance imaging was done with a Microsanj thermoreflectance image analyzer. Raman and photoluminescence measurements were carried out on a HORIBA LabRAM HR800 Raman spectrometer.

**Data availability.** The data that support the findings of the study are available from the corresponding author upon reasonable request.

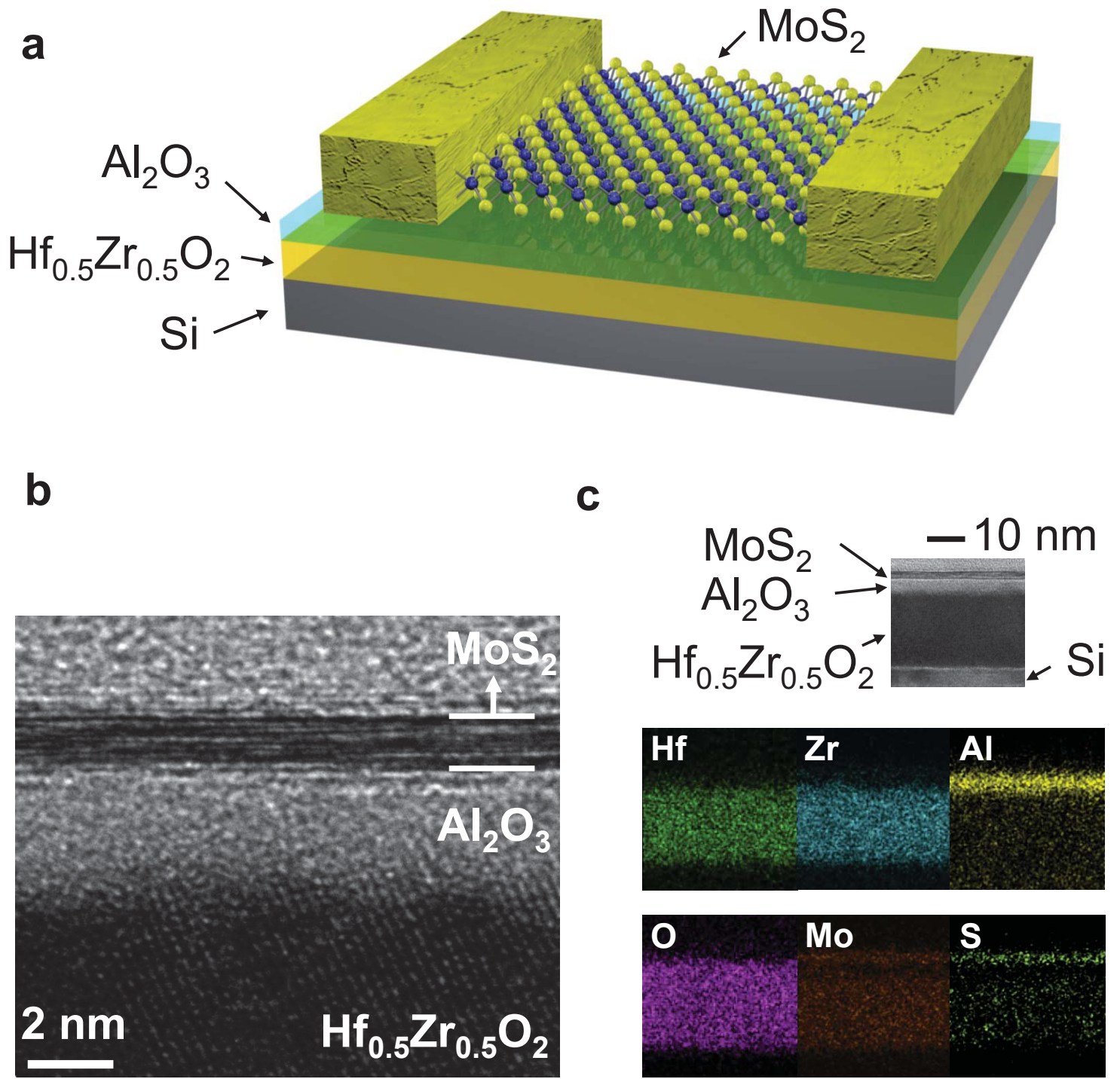

**Figure 1**

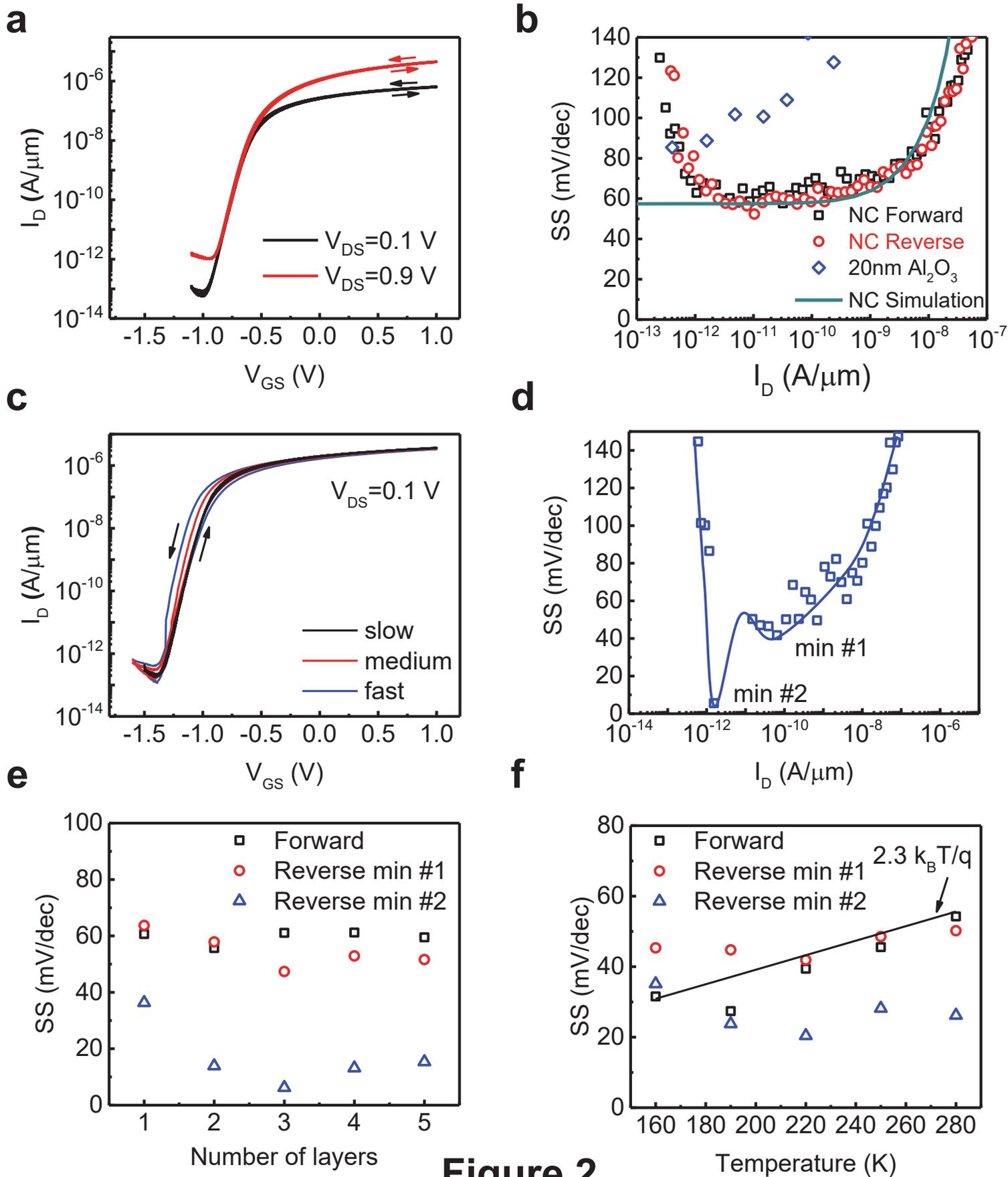

**Figure 2**

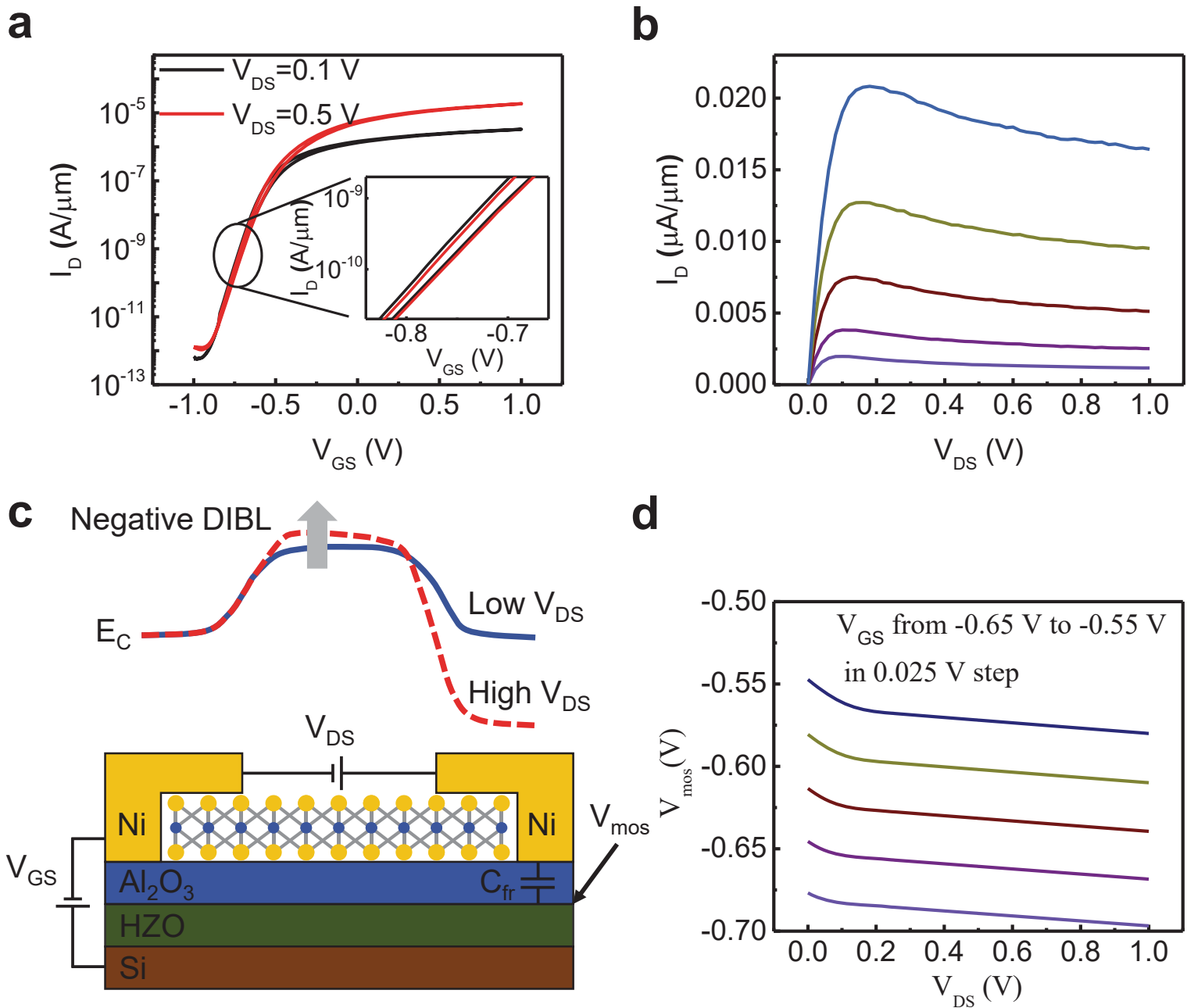

**Figure 3**

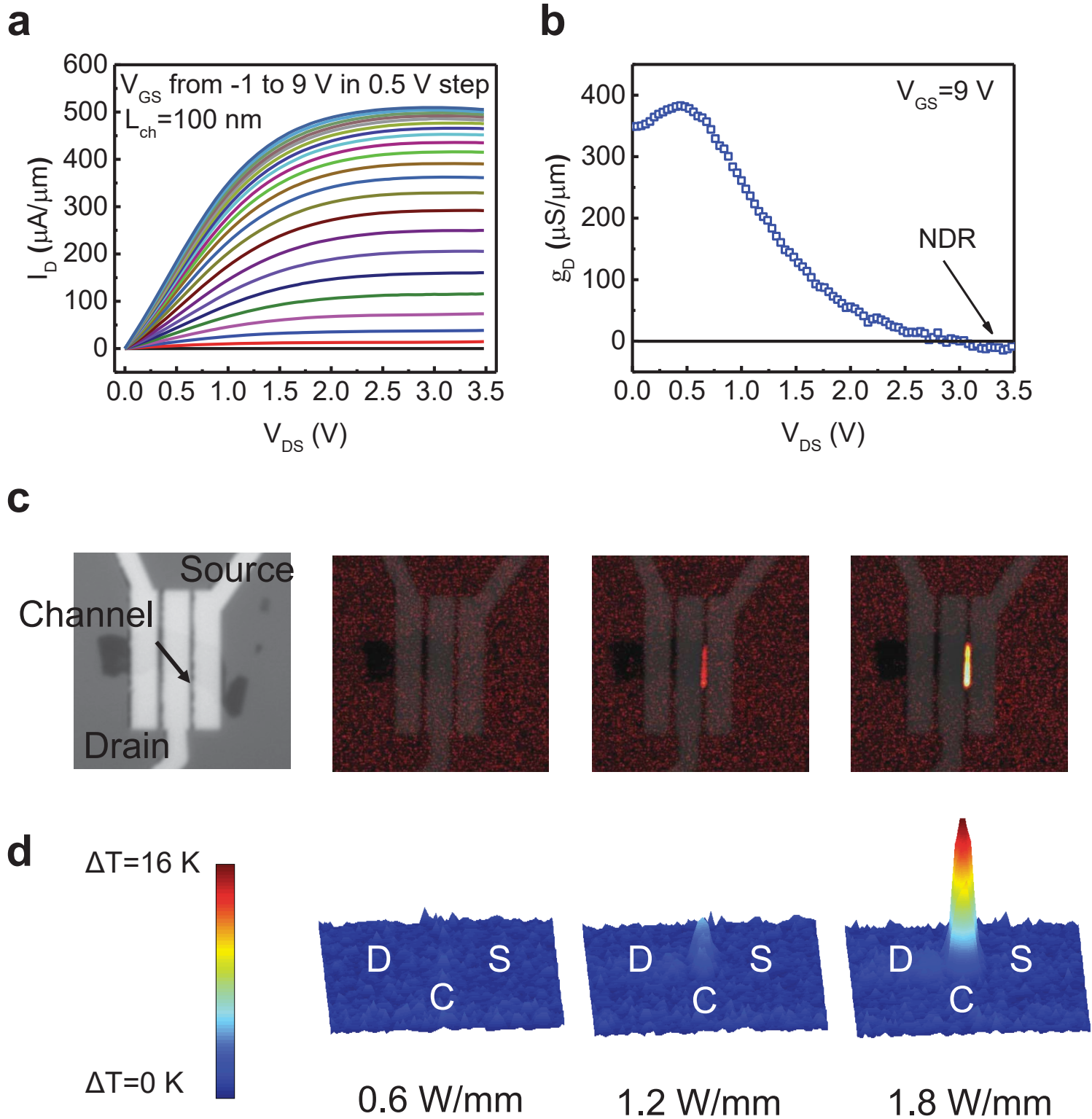

**Figure 4**

Supplementary Information for:

# Steep Slope Hysteresis-free Negative Capacitance MoS$_2$ Transistors

Mengwei Si[1,3], Chun-Jung Su[2], Chunsheng Jiang[1,4], Nathan J. Conrad[1,3], Hong Zhou[1,3], Kerry D. Maize[1,3], Gang Qiu[1,3], Chien-Ting Wu[2], Ali Shakouri[1,3], Muhammad A. Alam[1] and Peide D. Ye*[,1,3]

[1]*School of Electrical and Computer Engineering, Purdue University, West Lafayette, Indiana 47907, United States*

[2]*National Nano Device Laboratories, Hsinchu 300, Taiwan*

[3]*Birck Nanotechnology Center, Purdue University, West Lafayette, Indiana 47907, United States*

[4]*Tsinghua National Laboratory for Information Science and Technology, Institute of Microelectronics, Tsinghua University, Beijing 100084, China*

* Address correspondence to: yep@purdue.edu (P.D.Y.)

# 1. Fast I-V measurement of ferroelectric MOS capacitors

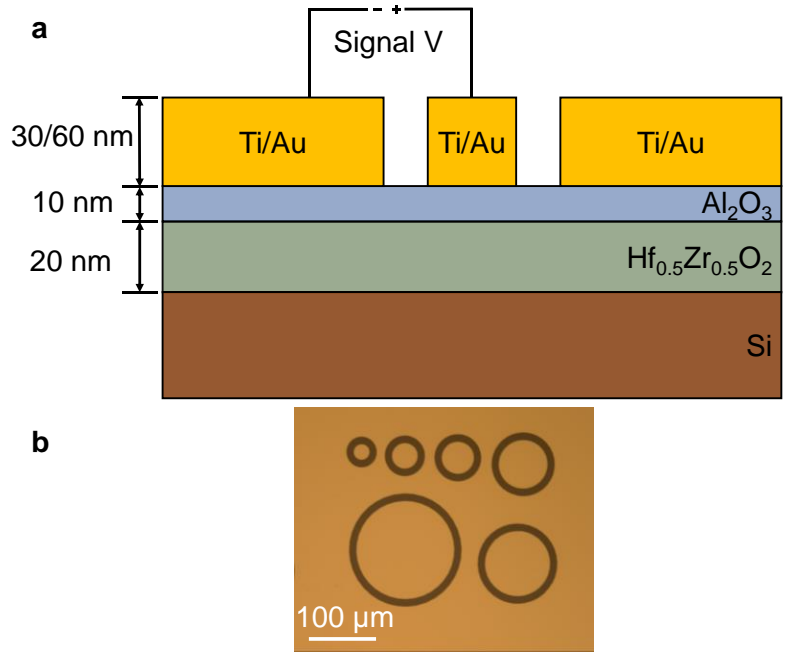

**Figure S1 | Illustration of MOS structure for Pulsed I-V measurement. a** Schematic diagram of a ferroelectric MOS capacitor for fast I-V measurement. **b** Optical image of the ferroelectric MOS capacitors from top view.

To study ferroelectric characteristics of the gate stack, a ferroelectric test structure is designed for fast I-V and C-V measurement. Fig. S1a shows the schematic of the ferroelectric MOS capacitor for test structure and Fig. S1b shows an optical image of the ferroelectric MOS capacitors. Hafnium zirconium oxide (HZO) was deposited by atomic layer deposition (ALD) for 20 nm as ferroelectric insulator layer on heavily doped silicon substrates. Another 10 nm aluminum oxide ($Al_2O_3$) layer was deposited as an encapsulation layer for capacitance matching and to prevent degradation of HZO due to air exposure. The annealing process was performed in rapid thermal annealing (RTA) in nitrogen ambient for 1 minute at various temperatures. Ti/Au with 30 nm/60 nm was used as electrode metal.

To validate the ferroelectricity of the gate stack used in this work, current response to a triangular voltage signal was measured to characterize the hysteresis loop of polarization versus electric field (P-E). All current responses from no RTA to 600 °C RTA deviate from a square wave signal, indicating the MOS capacitors measured in this work is not linear capacitors (Fig. S2a). The hysteresis loops of P-E at different temperatures are obtained from the integration of current response as a function of voltage, to obtain the polarized charge density[1]. From the hysteresis loop of P-E, it is confirmed that the samples with 400 °C and 500 °C exhibit stronger ferroelectricity, compared to those with no RTA or 600 °C.

Grazing incidence X-ray diffraction (GI-XRD) analysis in Fig. S2d depicts the crystallization behaviors of HZO with no RTA and 400 °C. The sample with 400 °C reveals apparent orthorhombic phases (o-phases). The formation of non-centrosymmetric o-phase is believed to lead to the ferroelectricity of HZO films after annealing[2,3], as confirmed in Fig. S2b. The slightly crystallized HZO found in the sample with no RTA is attributed to the thermal budget of ALD $Al_2O_3$ deposition.

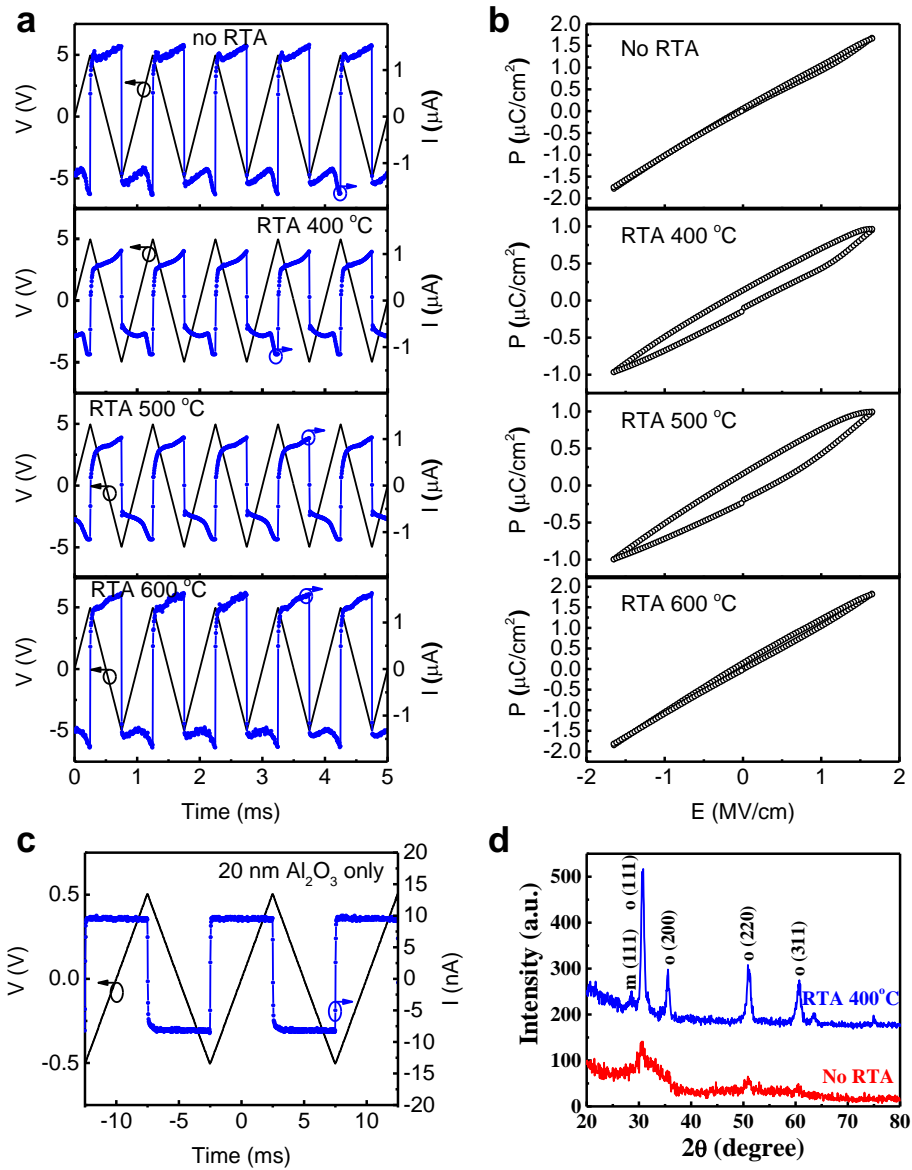

**Figure S2 | Ferroelectricity in the gate stack. a** Current response to a triangular voltage signal of the ferroelectric capacitor in Fig. S1 without RTA and with RTA from 400 ℃ to 600 ℃ in $N_2$ ambient for 1 min. **b** Temperature dependence of the P-E hysteresis curves obtained from a. **c** Current response of a linear capacitor with 20 nm $Al_2O_3$ only as dielectric. **d** GI-XRD diffractograms of HZO. The formation of non-centrosymmetric o-phase is believed to lead to the ferroelectricity of HZO films after annealing at 400℃.

## 2. Gate leakage current of MoS₂ NC-FETs

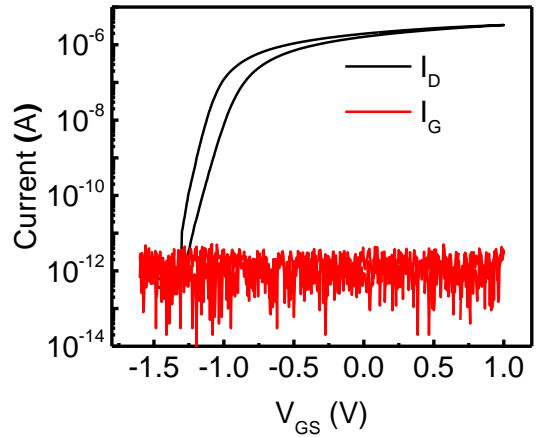

**Figure S3 | Gate leakage current in MoS₂ NC-FETs.** Gate leakage current and $I_D$-$V_{GS}$ characteristics simultaneously measured in the MoS₂ NC-FET for Fig. 2c.

The gate leakage current was measurement simultaneously with $I_D$, as shown in Fig. S3. It is the gate leakage current and $I_D$-$V_{GS}$ characteristics simultaneously measured in the MoS₂ NC-FET for Fig. 2c. A constant gate leakage current ~pA level means the gate leakage current is negligible in subthreshold region and the measured leakage is the lower detection limit of the equipment, as a medium power SMU is used for gate leakage current to speed up measurement here instead of a high-resolution SMU for $I_D$.

## 3. Effect of RTA temperature on the subthreshold slope of MoS₂ NC-FETs

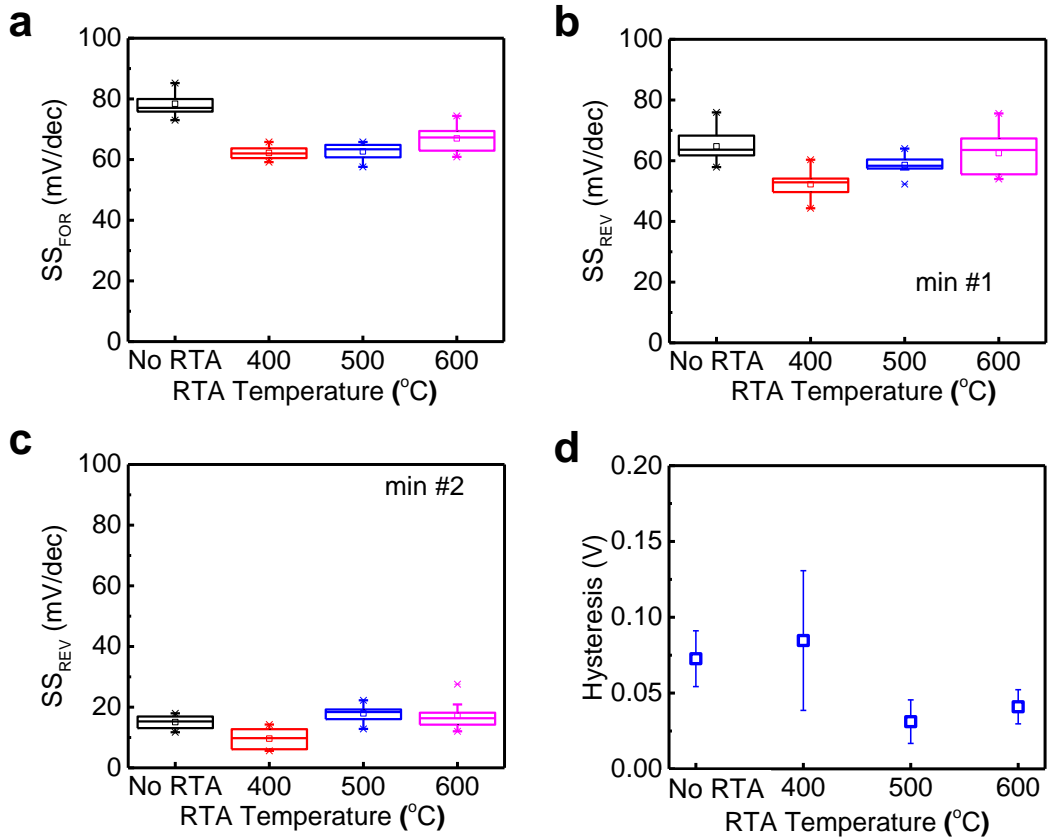

**Figure S4 | Statistic study of the effect of RTA temperature on the subthreshold slope and hysteresis of MoS₂ NC-FETs. a** $SS_{For}$, **b** $SS_{Rev,min\#1}$, **c** $SS_{Rev,min\#1}$. **d** hysteresis. Each data point contains the measurement of at least 8 individual devices with the same fabrication process. The SS and hysteresis presented here are all from $I_D$-$V_{GS}$ characteristics measured at 5 mV $V_{GS}$ step. The hysteresis is measured as $V_{GS}$-difference between forward and reverse sweeps at $I_D$=1 nA/μm and at $V_{DS}$=0.1 V. All the devices have the device structure as shown in Fig. 1.

As the annealing temperature is quite critical to ferroelectricity of the gate stack, we carried out the statistic study of the effect of RTA temperature on the SS of MoS₂ NC-FETs. As the RTA was performed after the gate stack deposition and before the transfer of MoS₂ flake, only the substrate, HZO and Al₂O₃ were affected. Fig. S4a-c shows the $SS_{For}$, $SS_{Rev,min\#1}$ and $SS_{Rev,min\#2}$ versus RTA temperature, respectively. It is found that devices with 400 °C RTA show the lowest

SS for all three SS characteristics. Meanwhile, devices with 500 °C RTA have lower $SS_{For}$ and $SS_{Rev,min\#1}$ than devices without RTA and devices with 600 °C RTA. This RTA temperature dependence of SS is very consistent with the results from Fig. S2. Devices with 400 °C or 500 °C RTA have lower SS comparing with devices without RTA or with 600 °C RTA because the stronger ferroelectricity, as shown in Fig. S2b. A statistic study on temperature dependent hysteresis is shown in Fig. S4d. It is found that $MoS_2$ NC-FETs with 500 °C RTA exhibit the lowest hysteresis comparing with devices without RTA, devices with RTA at 400 °C and 600 °C. All hysteresis data collected here is from $I_D$-$V_{GS}$ characteristics measured in fast sweep mode with 5 mV $V_{GS}$ step and at $V_{DS}$=0.1 V.

## 4. Layer number determination of $MoS_2$ flake and mono-layer $MoS_2$ NC-FET

Monolayer, bi-layer and multi-layer $MoS_2$ flakes were identified using three techniques: Raman shift[4], photoluminescence spectra[5] and AFM measurement[6]. There are two characteristic Raman modes, the in-plane vibrational mode and the out-of-plane vibrational mode with $\Delta\omega$=18.5 $cm^{-1}$ indicating mono-layer and $\Delta\omega$=21.4 $cm^{-1}$ indicating bi-layer, as shown in Fig. S5a. Meanwhile, mono-layer $MoS_2$ is well known to have a direct bandgap so that there is a strong peak in photoluminescence spectra as shown in Fig. S5b. It is straight forward to distinguish mono-layer $MoS_2$ from bi-layer or few-layer $MoS_2$. AFM measurement is also applied to determine the thickness and a mono-layer $MoS_2$ flake in this work is measured to be around 0.9 nm, as shown in Fig. S5c. Fig. S5d shows the $I_D$-$V_{GS}$ characteristics of a mono-layer $MoS_2$ NC-FET. Severe SS degradation is observed at low $V_{DS}$ due to the large Schottky barrier height for mono-layer $MoS_2$ at metal/channel contacts.

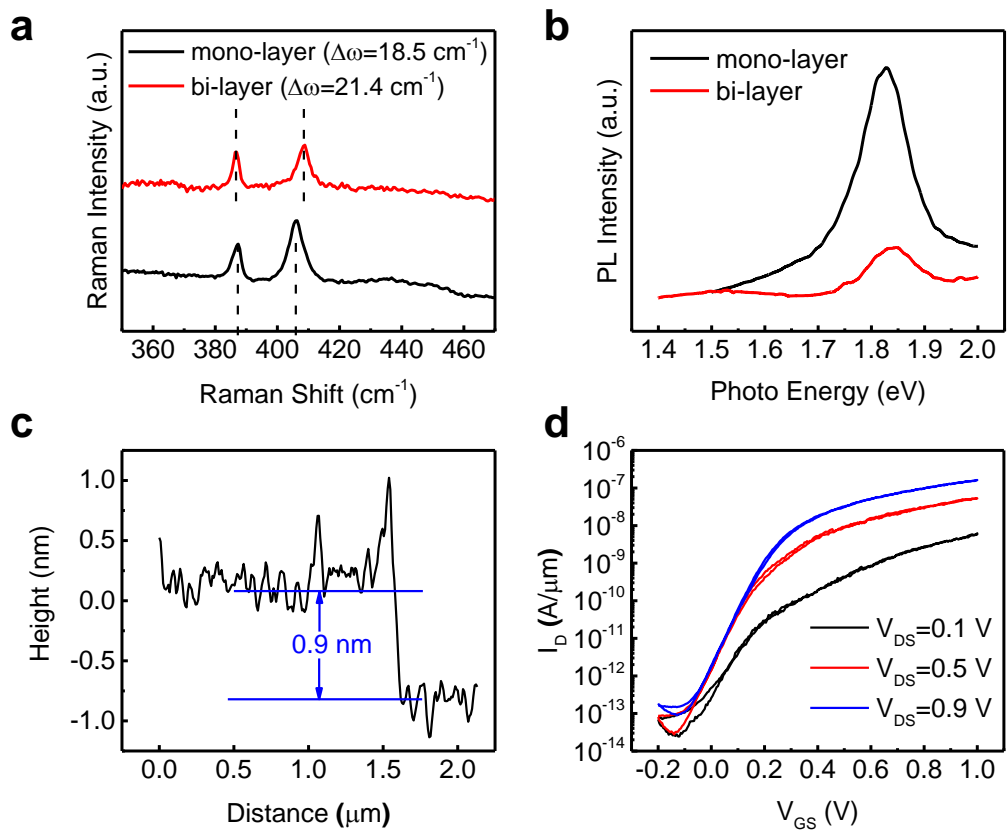

**Figure S5 | Mono-layer identification and monolayer MoS₂ NC-FET. a** Raman spectrum measurement of mono-layer and bi-layer MoS₂. **b** Photoluminescence measurement of single-layer and bi-layer MoS₂. **c** AFM measurement of a mono-layer MoS₂ flake. **d** $I_D$-$V_{GS}$ characteristics of a mono-layer MoS₂ NC-FET with 0.5 μm channel length.

## 5. Low temperature measurement of a bi-layer MoS₂ NC-FET

Fig. S6 shows the low temperature measurement of a bi-layer MoS$_2$ NC-FET from 160 K to 220 K. The device has a channel length of 0.5 μm and a channel width of 2.5 μm. The low temperature electrical data was collected with a Lakeshore TTP4 probe station. Measured SS is below the thermionic limit down to 220 K. SS below 190 K shows above the thermionic limit because of stronger impact of Schottky barrier on SS.

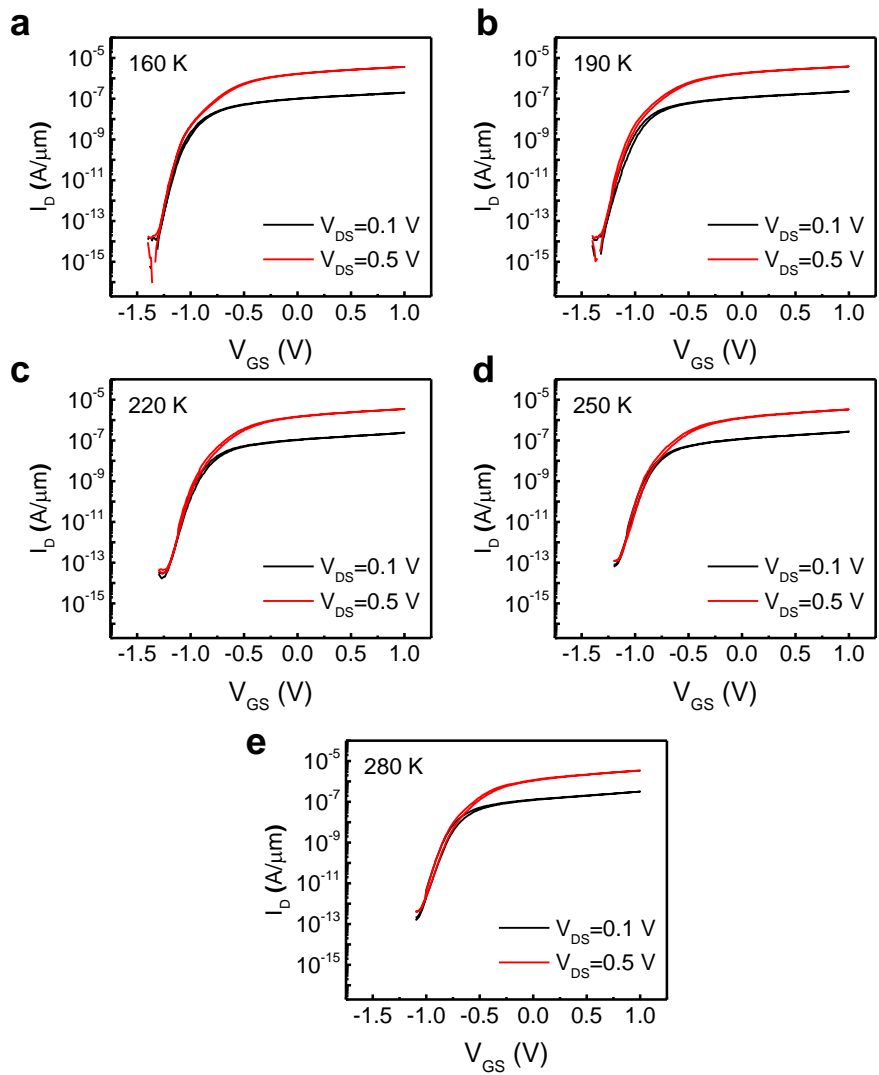

**Figure S6 | Low temperature measurement of a bi-layer MoS₂ NC-FET.** I$_D$-V$_{GS}$ characteristics of a bi-layer MoS$_2$ NC-FET with 0.5 μm channel length, 2.5 μm channel width. **a** 160 K. **b** 190 K. **c** 220 K. **d** 250 K. **e** 280 K.

## 6. Experiment setup for thermoreflectance imaging

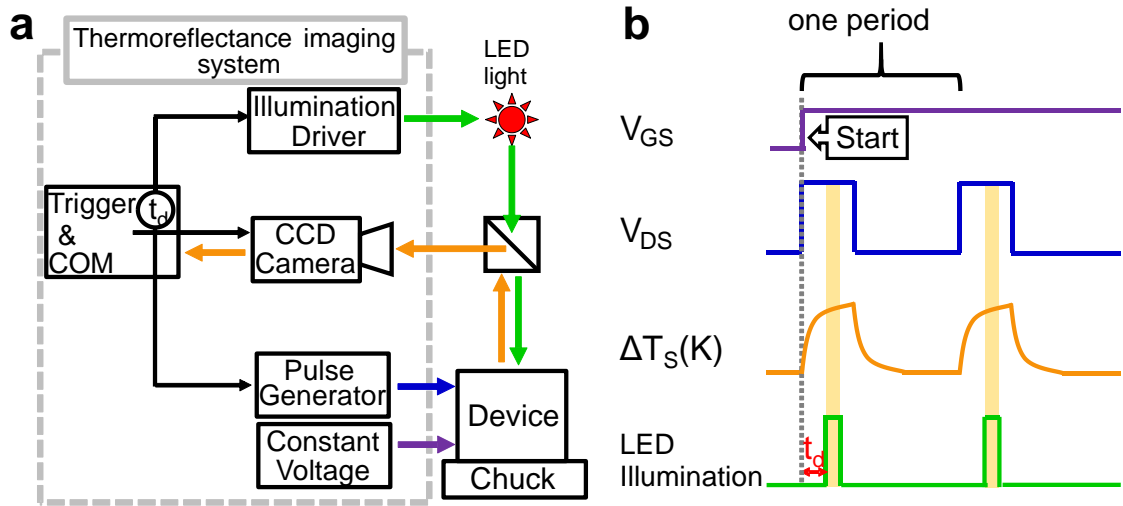

**Figure S7 | Illustration of thermoreflectance imaging measurement system. a** A schematic of thermoreflectance imaging system. A pulse generator ($V_{DS}$) and a constant voltage source ($V_{GS}$) drive the transistor. A control computer triggers the illumination driver and the CCD camera for a given delay time with respect to $V_{DS}$. **b** Timing diagram for transient TR imaging with a given LED delay time ($t_d$).

The thermoreflectance (TR) measurement system setup is shown in Fig. S7[7,8]. A high-speed LED pulse illuminates the device, and a synchronized chare coupled device (CCD) camera captures the reflected image, as shown in Fig. 7a. The MoS$_2$ is illuminated through an LED ($\lambda$ = 530 nm) via an objective lens. The reflected light from the surface of MoS$_2$ channel is captured on a variable frame rate, 14-bit digitization, Andor CCD camera.

For the transient measurement of temperature, the device is periodically turned ON and OFF by a $V_{DS}$ pulse, as shown in Fig. S7b, allowing the channel to heat and cool, respectively. By controlling the delay of the LED pulse with respect to the beginning of the $V_{DS}$ pulse, the TR image can capture different phases of the transient heating and cooling kinetics. The delay time for the LED illumination can be varied and each illumination pulse acts as a camera shutter. Every $V_{DS}$ cycle produces an image capturing the thermal state of the substrate at a given time delay. The average of these images improves the signal-to-noise ratio and produces a high-resolution map of temperature. In this work, temperature was measured at the last 100 μs of the 1 ms drain voltage pulse (10 ms period).

The change in reflectivity ($\Delta R$) under visible spectral range is proportional to the change in temperature, so that once the TR coefficient is obtained, $\Delta R$ can be mapped to differential increase in temperature ($\Delta T_S$). Unfortunately, TR coefficient must be calibrated, because it depends on the wavelength, the angle of incidence, and the polarization of the incident light, as well as the surface properties of the reflecting material. The calibration is performed by heating the sample by placing it on an external microthermoelectric stage. The temperature of the sample is monitored by micro-thermocouple while capturing the reflection changes by the CCD camera. The TR coefficient for the specific setup is obtained by plotting the change in reflectivity as a function of temperature measured by the thermocouple. Here, TR coefficient is calibrated on exfoliated $MoS_2$ flakes.

## 7. Simulation of MoS₂ NC-FETs

As shown in Fig. S8a a negative capacitance MoS$_2$ transistor can be treated as an intrinsic MoS$_2$ transistor in series with an HZO ferroelectric capacitor. In addition, the electrical behavior of HZO ferroelectric capacitor can be described by Landau-Khalatnikov (LK) equation[9-11]. Landau coefficients are extracted from the experimental P-E curve of HZO. For the intrinsic MoS$_2$ transistor, one can obtain its transfer characteristic and output characteristic using classical drift-diffusion method. To simulate the experimental device (metal (Heavily-doped silicon)-ferroelectric oxide-insulator-semiconductor), we will assume that the potential distribution is essentially uniform across the gate dielectric, which simplifies the overall analysis by allowing one to decouple the HZO dielectric from the standard MOSFET structure[12-14]. In fact, the errors caused by this approximation can be ignored when the thickness of ferroelectric layer is not too thick[15,16]. The other device parameters are extracted from the experimental transfer characteristics. All simulations assume 1 μm channel length, 8.6 nm thick MoS$_2$ flake, and 2 nm Al$_2$O$_3$ capping, unless otherwise specified.

Landau coefficients ($\alpha$, $\beta$, $\gamma$) are extracted from the P-E measurement on the TiN/HZO/TiN structure, as shown in Fig. S8c, in which ALD HZO process condition is exactly same as the one for the HZO/Al$_2$O$_3$ stacks but with TiN as top and bottom metallic electrodes. The complete LK equation is written as[17],

$$V_{GS} = V_{mos} + V_f = V_{mos} + 2t_f \alpha Q_{av} + 4t_f \beta Q_{av}^3 + 6t_f \gamma Q_{av}^5 + \rho t_f \frac{dQ_{av}}{dt} \qquad (1)$$

$$Q_{av} = \frac{Q_{ch} + Q_{p1} + Q_{p2}}{WL} \qquad (2)$$

$$Q_{p1} = C_{fr} W_{ch}(V_{mos} - V_S) \qquad (3)$$

$$Q_{p2} = C_{fr} W_{ch}(V_{mos} - V_D) \qquad (4)$$

where $Q_{av}$ is the average gate charges density per area. $Q_{ch}$ is the intrinsic channel area charge, $Q_{p1}$ is the parasitic charges caused by the source-gate capacitance, and $Q_{p2}$ is the parasitic charges caused by the drain-gate capacitance. $\alpha$, $\beta$, and $\gamma$ are Landau coefficients, which are material dependent constants; $t_f$ is the thickness of the ferroelectric film; and $V_f$ is the external applied voltage across the ferroelectric layer. $\rho$ is an equivalent damping constant of HZO.

The Landau coefficients are extracted to be $\alpha$=-1.1911e8 m/F, $\beta$=4.32e9 $m^5$/F/$coul^2$, and $\gamma$=0 $m^9$/F/$coul^4$, as shown in Fig. S8c. Fig. S8d shows the simulation results based on these experimental Landau coefficients which exactly match with our experimental results. Based on the Landau coefficients extracted from experimental P-E and eqn. (1), the capacitance of ferroelectric capacitor ($C_{FE}$) can be calculated using experimental Landau coefficients,

$$C_{FE} = \frac{dQ_{av}}{dV_f} = \frac{1}{2\alpha t_f + 12\beta t_f Q_{av}^2 + 30\gamma t_f Q_{av}^4} \tag{5}$$

The internal gain condition and the non-hysteretic condition for MoS$_2$ NC-FETs are discussed based on the experimental P-E and extracted Landau coefficients. To prevent hysteretic behavior and obtain a steep SS at the same time, some design rules must be obeyed. These design principles could be derived from its small-signal capacitance circuit of a 2D NC-FET as shown in Fig. S8b. SS can be written as,

$$SS = \frac{2.3k_BT}{q} \cdot \frac{1}{\frac{\partial \phi_s}{\partial V_{gs}}} = \frac{2.3k_BT}{q}\left(1 + \frac{C_{2D}}{C_{ox}}\right) \cdot (1 - \frac{C_{device}}{|C_{FE}|}) \tag{6}$$

$$C_{device} = 2C_{fr} + \frac{C_{2D}C_{ox}}{C_{2D}+C_{ox}} \tag{7}$$

Note that $C_{fr}$ is the parasitic capacitance. SS must satisfy the condition, 0<SS<2.3$k_B$T/q, so that non-hysteretic behavior and a sub-thermionic SS (internal gain>1) could be obtained at the same time. The constraint conditions as the equations (8, 9) deduced from (6) are,

$$C_{device} < |C_{FE}| \tag{8}$$

$$|C_{FE}| < C_{eq} \tag{9}$$

$$C_{eq} = \left(1 + \frac{C_{ox}}{C_{2D}}\right) \cdot C_{device} \tag{10}$$

If no parasitic capacitance is considered as $C_{fr}$=0, the constraint conditions and SS become,

$$\frac{C_{2D}C_{ox}}{C_{2D}+C_{ox}} < |C_{FE}| \tag{11}$$

$$|C_{FE}| < C_{ox} \tag{12}$$

$$SS = \frac{2.3k_BT}{q}\left(1 + \frac{C_{2D}(|C_{FE}|-C_{ox})}{|C_{FE}|C_{ox}}\right) \tag{13}$$

To satisfy non-hysteretic conditions, $|C_{FE}|$ need to be greater than $C_{device}$ (eqn. (8)), while to satisfy internal gain condition (internal gain>1, SS<2.3$k_B$T/q), $|C_{FE}|$ need to be less than $C_{eq}$ (eqn. (9)). Note that $C_{eq}$ equals to $C_{ox}$ if $C_{fr}$=0. $C^{-1}$ of $|C_{FE}|$, $C_{device}$, and $C_{eq}$ are compared as shown in Fig. S8e with different $t_f$. It is clear to see that if $t_f$ is greater than 72.5 nm, $|C_{FE}|$ becomes smaller than $C_{device}$ which is against eqn. (8) so that hysteresis will be introduced, as shown in Fig. S8f. If $|C_{FE}|$ is less than $C_{eq}$, the design satisfies the internal gain condition where SS can be less than 2.3 $k_B$T/q, as shown in Fig. S8e. When the gate voltage is in subthreshold region, $|C_{FE}|$ is less than $C_{eq}$ among all $t_f$.

The internal gain condition and non-hysteresis condition are directly related with the $C_{fr}$. If $C_{fr}$=0, the internal gain condition ($|C_{FE}|<C_{ox}$) as eqn. (12), can't be fulfilled since the minimum $|C_{FE}|$ obtained for 20 nm HZO from eqn. (2) is about $|C_{FE}|$=13.1 $\mu$F/cm$^2$, which is larger than the $C_{ox}$=3.54 $\mu$F/cm$^2$ (2 nm Al$_2$O$_3$). Therefore, $C_{fr}$ must be considered to fulfill the internal gain conditions, as calculated in Fig. S8e. With the existence of $C_{fr}$, $|C_{FE}|$ can be smaller than $C_{eq}$, which fulfills the internal gain condition in eqn. (9). Fig. S8g shows the impact of $C_{fr}$ on the SS vs. $I_D$ characteristics. It is clear that if $C_{fr}$=0, the SS of the MoS$_2$ NC-FET is the same as 2.3$k_B$T/q so that no internal gain can be obtained as predicted by eqns. (12, 13). However, if we consider the impact of $C_{fr}$, SS can be less than $2.3k_BT/q$ (internal gain>1) because eqn. (9) is fulfilled as shown in Fig. 8e.

Fig. 8h shows the $t_{ox}$-$t_f$ design plane of the device. The boundary line between two regions represents the capacitance match: $-C_{FE}=C_{device}$. The cyan area represents the design space of transfer characteristics with non-hysteresis and a steep SS. Even though the subthreshold slope would be reduced when $t_f$ increases, the hysteresis must be avoided in logic applications. Thus, the device geometries ($t_f$-$t_{ox}$) should be co-optimized to avoid the hysteresis and achieve a steep SS at the same time.

The simulation results MoS$_2$ NC-FETs are discussed in details after satisfying the internal gain and non-hysteretic conditions. As shown in Fig. S9a, it can be observed that I$_{DS}$ decreases obviously as $t_f$ increases for a given gate voltage when the device works in the depleted regime (V$_{GS}$<V$_{FB}$). V$_{FB}$ is defined as the gate voltage when the total gate (or channel) charges reaches zero. In a junctionless transistor, this critical voltage differentiates between depletion-mode subthreshold operation vs. accumulation mode above threshold operation[15]. Note that V$_{FB}$ is bigger than V$_{FB0}$ (flat-band voltage when V$_{DS}$=0 V) because there is a depleted region in the drain terminal when V$_{DS}$ is not zero. Thus, the increasing of $t_f$ lowers the off-state current significantly and improve threshold voltage compared with its conventional MoS$_2$ transistor (when $t_f$=0 nm, a MoS$_2$ NC-FET is reduced to a MoS$_2$ transistor). In contrast, in the on-state accumulation regime (V$_{GS}$>V$_{FB}$), I$_{DS}$ increases when $t_f$ increases. In other words, both on and off state performances improve with $t_f$, so long the transistor is operated in the NC-FET mode. The phenomenon can be explained as follows. Fig. S9b shows that the interfacial potential (V$_{mos}$) varying with V$_{GS}$ for different $t_f$. When V$_{GS}$ is smaller than V$_{FB}$ (in the depleted regime), V$_{mos}$ deceases with $t_f$ increasing while when V$_{GS}$ is bigger than V$_{FB}$ (in the accumulation regime), V$_{mos}$ increase with $t_f$ increasing.

Thus, the off-state current can be lowered and on-state current can be improved at the same time. Among the range of HZO thicknesses possible, $t_f$=20nm was chosen for processing convenience.

For drift-diffusion based transistors, the subthreshold slope can be estimated as 2.3 $k_BT/(d\varphi_S/dV_{GS})$. For negative capacitance FETs, the DC voltage gain (defined as the body factor $m=d\varphi_S/dV_{GS}$) can be larger than 1, so that SS<2.3 $k_BT$ in this case. Fig. S9c shows that $m$ varies with $V_{GS}$ for different $t_f$. It can be seen that m>1 in the subthreshold regime for a MoS$_2$ NC-FET and m enlarges when $t_f$ increases for a given $V_{GS}$. It causes that SS can be smaller 60 mV/dec in a big range of $I_{DS}$ as shown in Fig. S9d. The results from our analytical model match well with those from the experimental data, as shown in Fig. 2.

Fig. S9e shows the transfer characteristics of a MoS$_2$ NC-FET for different $V_{DS}$. Contrary to the normal MOSFETs, there is a reverse DIBL effect in the transfer characteristics of the MoS$_2$ NC-FET. That is, the threshold voltage increases when $V_{DS}$ increases. In order to understand this unique property, the $V_{mos}$ varying with $V_{GS}$ for different $V_{DS}$ is shown in Fig. S9f. One observes that $V_{mos}$ reduces when $V_{DS}$ increases in the subthreshold voltage. On the other hand, $V_{DS}$ has almost no impact on $I_{DS}$ of the intrinsic MoS$_2$ transistor as shown in Fig. S9g. The reason is that while the DIBL effect of a long-channel intrinsic MoS$_2$ transistor can be neglected, but this is not true for MoS$_2$ NC-FET where $I_{DS}$ is reduced with increasing $V_{DS}$.

The NC-FET also exhibits a characteristic negative differential resistance (NDR) in the output characteristics. Fig. S9h illustrate the output characteristics of a MoS$_2$ NC-FET for different $V_{GS}$ (with $t_f$=20 nm). There is a clear NDR effect when the device works in the saturation region ($V_{DS}$>$V_{GS}$-$V_{th}$). Simulated $V_{mos}$ vs $V_{DS}$ curves for different $V_{GS}$ are shown in Fig. 3d. It is seen that $V_{mos}$ decreases when $V_{DS}$ increases when the device works in the saturation region. On the other hand, $V_{DS}$ has a small impact on $I_{DS}$ of the intrinsic MoS$_2$ transistor when the device works in the saturation region. Thus, $V_{mos}$ dominates the saturation current of the MoS$_2$ NC-FET. That is, the saturation current is reduced with increasing $V_{DS}$.

Although the non-hysteretic conditions have been achieved in steady-state, hysteresis during $I_D$-$V_{GS}$ measurements can still appear as the result of dynamic dumping factor ρ>0. Because the steady-state model is ideal while the actual measurement process is dynamic because the rise time of the gate voltage cannot be infinity so that $V_{mos}$ cannot follow the change speed of $V_{GS}$, which leads to the hysteresis (Fig. S10a). If there is no damping constant, as shown in Fig. S10d and S10e, no hysteresis can be observed for a MoS$_2$ NC-FET with 20 nm HZO. But if we add a dumping resistor ($R_{FE}$ in Fig. S8b) so that ρ is greater than zero, hysteresis will exist again, as shown in Fig. S10b and S10c. Thus, the second origin of hysteresis is the existence of dumping constant in the ferroelectric HZO.

Based on the discuss above, the hysteresis measured in this work is mostly dumping constant induced hysteresis, as shown in Fig. 2c in manuscript, which is measurement speed dependent. Therefore, our devices fulfill the condition of DC non-hysteretic and internal gain conditions. Meanwhile, by comparing the simulation results on parasitic capacitance, it can be concluded that the damping constant is the origin of the hysteresis and the parasitic capacitance causes the negative DIBL effect, as shown in Fig. S10b-e. And our experimental results in Fig. 3a in the manuscript qualitatively match with simulation results in Fig. S10. The experimental measured dumping factor is ρ ~ 30 Ωm for ferroelectric HZO[18], which is used in this work for the prediction of working speed for MoS$_2$ NC-FETs shown in Fig S10f. It can be seen that the MoS$_2$ NC-FETs still maintain decent hysteresis up to 0.1-1 MHz.

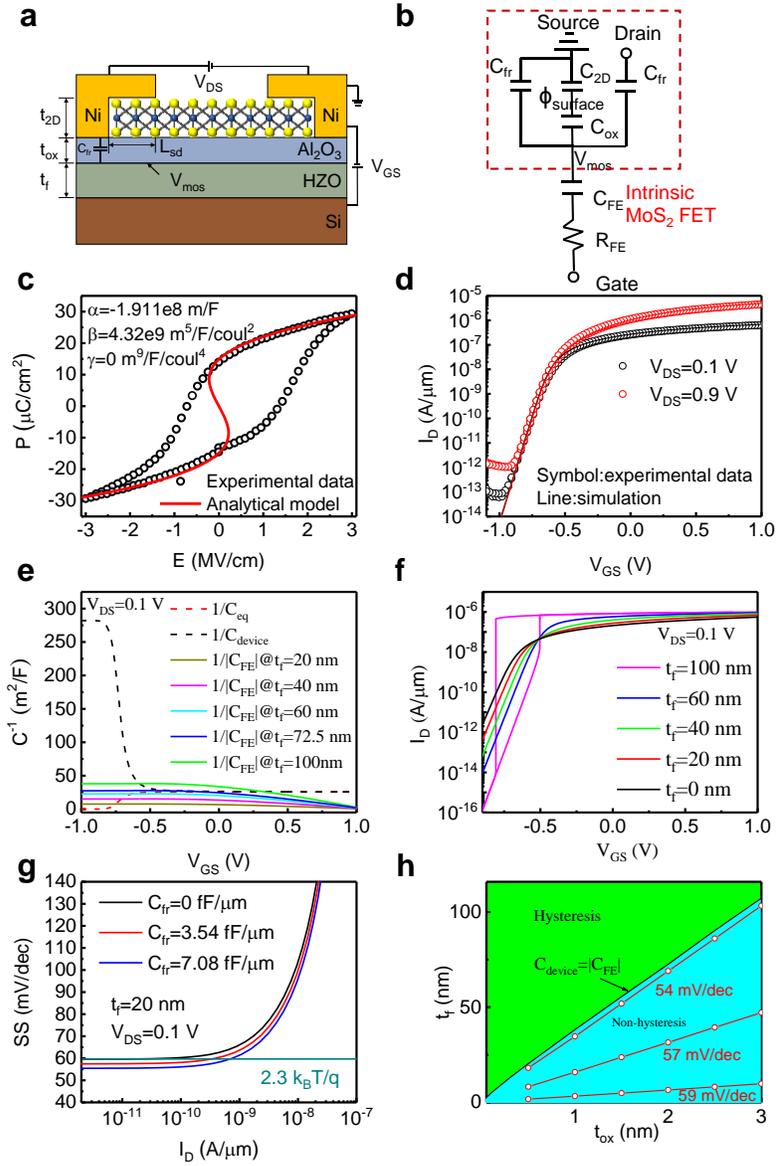

**Figure S8 | Experiments and simulations of the internal gain and non-hysteretic conditions on MoS₂ NC-FETs.** **a** Cross-section view of the MoS₂ NC-FET in simulation. **b** Simplified small-signal capacitance representation of a MoS₂ NC-FET for steady-state and dynamic simulation. $C_{2D}$ is the capacitance of MoS₂ channel, $C_{ox}$ is the capacitance of the $Al_2O_3$ layer, and $C_{FE}$ is the capacitance of HZO layer. **c** Experimental polarization-voltage measurement on ferroelectric HZO with MIM structure (TiN/HZO/TiN). **d** $I_D$-$V_{GS}$ characteristics for MoS₂ NC-FET as in Fig. 2a and the simulation based on parameters extracted from Fig. S8c. **c** Comparison of $C^{-1}$ between $C_{eq}$, $C_{device}$ and $|C_{FE}|$, which shows $|C_{FE}|>C_{device}$ to fulfill non-hysteretic condition and $|C_{FE}|<C_{eq}$ to fulfill internal gain condition. **d** $I_D$-$V_{GS}$ characteristics at $V_{DS}$=0.1 V for HZO films with various thicknesses. $|C_{FE}|<C_{device}$ at 100 nm HZO leads to a large hysteresis in steady-state. **e** SS vs. $I_D$ characteristics at different $C_{fr}$. **f** The $t_{ox}$-$t_f$ design plane of the MoS₂ NC-FET. The boundary line represents the capacitance match: $-C_{FE}$=$C_{device}$.

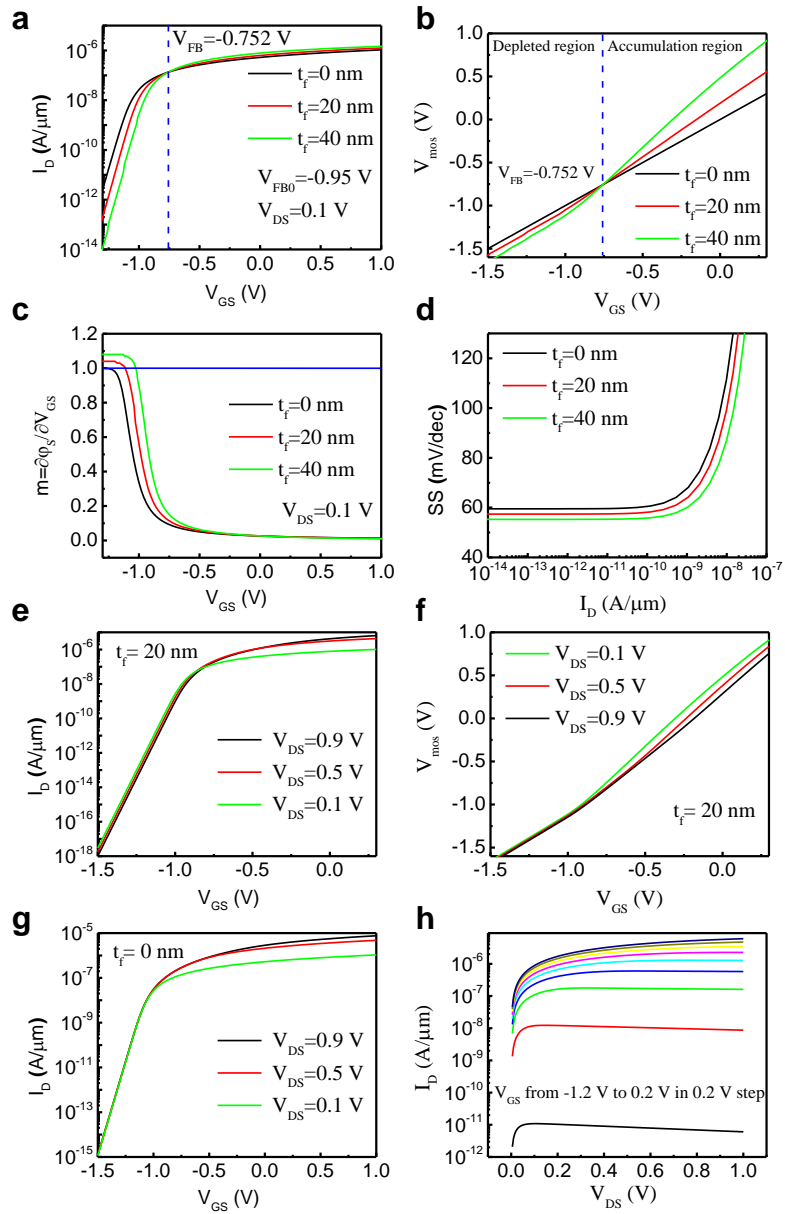

**Figure S9 | Simulation of negative DIBL and NDR effect on MoS₂ NC-FETs. a** $I_D$-$V_{GS}$ characteristics of MoS₂ NC-FETs with HZO thickness from 0 nm to 40 nm. **b** Interfacial potential vs. $V_{GS}$ with HZO thickness from 0 nm to 40 nm. **c** DC voltage gain of MoS₂ NC-FETs with HZO thickness from 0 nm to 40 nm. **d** SS-$I_D$ characteristics of MoS₂ NC-FETs with HZO thickness from 0 nm to 40 nm. **e** $I_D$-$V_{GS}$ characteristics of MoS₂ NC-FETs at different $V_{DS}$. **f** Interfacial potential vs. $V_{GS}$ of the same MoS₂ NC-FET at different $V_{DS}$. **g** $I_D$-$V_{GS}$ characteristics of MoS₂ FETs with no HZO dielectrics at different $V_{DS}$. **h** $I_D$-$V_{DS}$ characteristics of MoS₂ NC-FETs at different $V_{GS}$. Clear NDR can be observed at low $V_{GS}$.

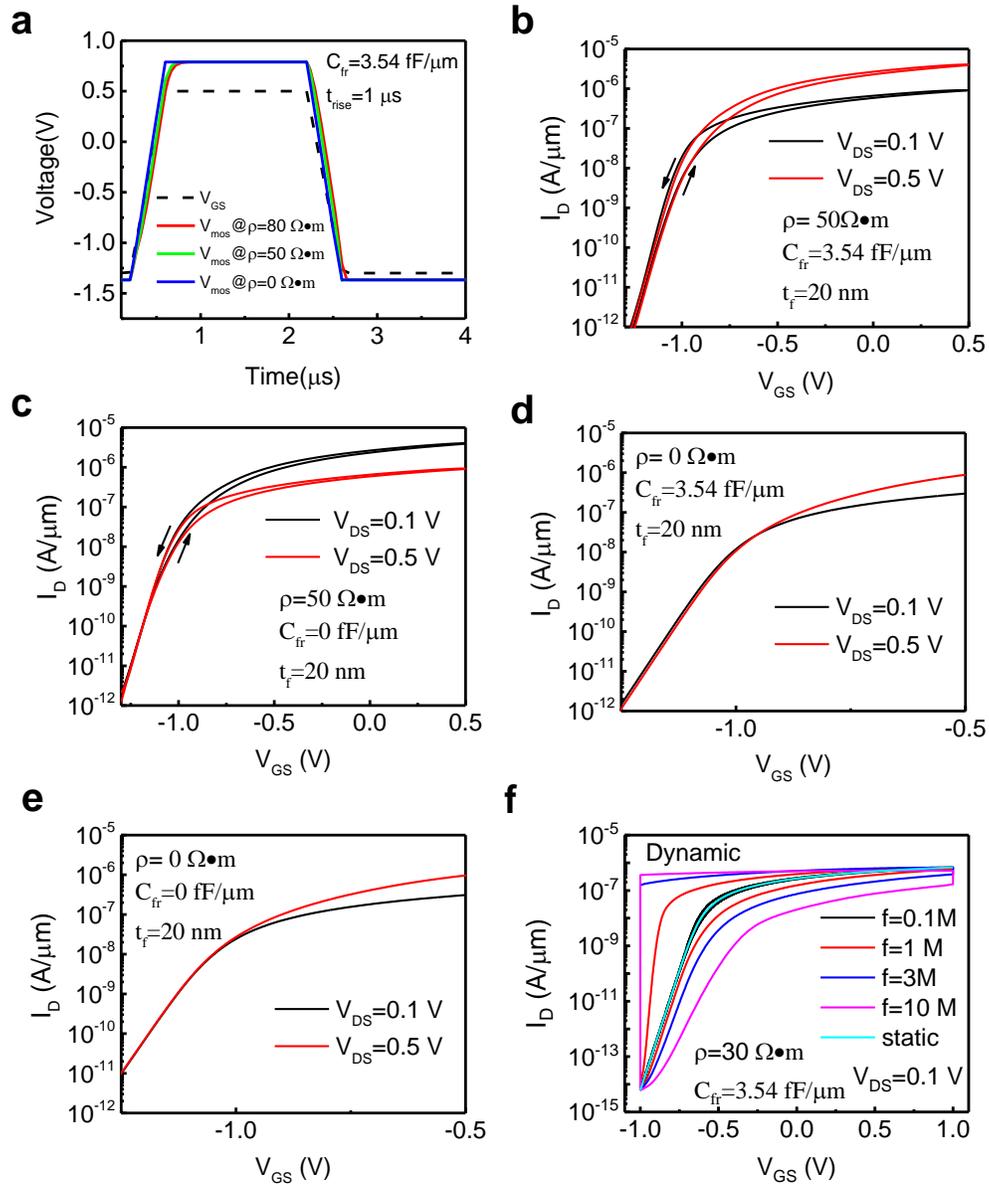

**Figure S10 | Simulation on stability and the effects of parasitic capacitance and dumping constant. a** Simulated transient behavior of a MoS$_2$ NC-FET. V$_{mos}$ cannot follow the change of V$_{GS}$, which leads to the hysteresis. **b** I$_D$-V$_{GS}$ characteristics with damping constant and parasitic capacitance for different V$_{DS}$. **c** I$_D$-V$_{GS}$ characteristics with damping constant and without the parasitic capacitance for different V$_{DS}$. **d** I$_D$-V$_{GS}$ characteristics without damping constant and with the parasitic capacitance for different V$_{DS}$. **e** I$_D$-V$_{GS}$ characteristics without damping constant and without the parasitic capacitance for different V$_{DS}$. **f** I$_D$-V$_{GS}$ characteristics for a MoS$_2$ NC-FET at different frequencies.